\documentclass[12pt]{article}

\newcommand{\x}{arXiv:}
\newcommand{\m}{\mathrm}
\newcommand{\be}{\begin{equation}}
\newcommand{\ee}{\end{equation}}
\newcommand{\ba}{\begin{eqnarray}}
\newcommand{\ea}{\end{eqnarray}}

\usepackage{graphicx}
\usepackage{amssymb}
\usepackage{amsmath}
\usepackage[T1]{fontenc} 
\usepackage[ansinew]{inputenc} 
\usepackage[nosort]{cite}
\newcommand{\inbar}{\vrule height1.57ex width.4pt depth0pt}
\newcommand{\SW}{\relax{\hbox{$\ \inbar\kern-.285em{\rm S}$}}}

\begin{document}
\thispagestyle{empty}
\begin{center}

\null \vskip-1truecm \vskip2truecm

{\Large{\bf \textsf{Holography of the QGP Reynolds Number}}}

{\large{\bf \textsf{}}}

{\large{\bf \textsf{}}}

\vskip1truecm

{\large \textsf{Brett McInnes
}}

\vskip0.1truecm

\textsf{\\ National
  University of Singapore}
  \vskip1.2truecm
\textsf{email: matmcinn@nus.edu.sg}\\

\end{center}
\vskip1truecm \centerline{\textsf{ABSTRACT}} \baselineskip=15pt
\medskip

The viscosity of the Quark-Gluon Plasma (QGP) is usually described holographically by the entropy-normalized dynamic viscosity $\eta/s$. However, other measures of viscosity, such as the kinematic viscosity $\nu$ and the Reynolds number $Re$, are often useful, and they too should be investigated from a holographic point of view. We show that a simple model of this kind puts an upper bound on $Re$ for nearly central collisions at a given temperature; this upper bound is in very good agreement with the observational lower bound (from the RHIC facility). Furthermore, in a holographic approach using only Einstein gravity, $\eta/s$ does not respond to variations of other physical parameters, while $\nu$ and $Re$ can do so. In particular, it is known that the magnetic fields arising in peripheral heavy-ion collisions vary strongly with the impact parameter $b$, and we find that the holographic model predicts that $\nu$ and $Re$ can also be expected to vary substantially with the magnetic field and therefore with $b$.

\newpage
\addtocounter{section}{1}
\section* {\large{\textsf{1. Two Measures of the QGP Viscosity}}}
The key result in holography \cite{kn:nat,kn:veron}, as applied to the Quark-Gluon Plasma (QGP) \cite{kn:youngman,kn:gubser,kn:janik,kn:chessch}, is the Kovtun-Son-Starinets \cite{kn:KSS,kn:SS,kn:sera1} computation of the ratio of the boundary field theory shear (or dynamic) viscosity $\eta$ to its entropy density $s$: what one might call the entropy-normalized dynamic viscosity, denoted henceforth by $\eta_s$, is given in natural units\footnote{$\eta_s$ is dimensionless in natural units. In SI units it has units of kelvin$\,\cdot\,$seconds, and the KSS relation takes the form $\eta_s \;=\; {\hbar \over 4\pi k_B} \approx 6.08 \times 10^{-13}$ K$\,\cdot\,$s \cite{kn:KSS}.} by
\begin{equation}\label{A}
\eta_s \; \equiv \; {\eta \over s} \;=\; {1 \over 4\pi}.
\end{equation}
This relation is computed at arbitrarily large coupling; at finite but large coupling one expects (with various subsequently discovered exceptions, some of which will be mentioned below) that this equality becomes an inequality, with $1/4\pi$ as the lower bound; there is reason to hope that the rate of variation upwards is not large, so the actual plasma should exhibit a value not too far above this.

Part of the reason for the attention generated by this result is that the actual QGP has indeed a value for $\eta_s$ which is surprisingly small, that is, not very far above $1/4\pi$ \cite{kn:song}; another reason is that (\ref{A}) is valid for a remarkable range of boundary field theories, in fact for all those which are isotropic and dual to Einstein gravity in the bulk. This \emph{universality} is important for a range of reasons, discussed in very clear detail in \cite{kn:nat}.

However, while it is true that, in many contexts, the viscosity only appears through $\eta_s$, in many other physical applications one is also interested in (some measure of) \emph{the viscosity itself} ---$\,$ and this includes discussions of the QGP, see for example \cite{kn:liaokoch,kn:facets}. Whether the actual viscosity of the QGP is ``small'' is a somewhat delicate question.

In classical hydrodynamics, there are two distinct measures of viscosity: the dynamic or shear viscosity $\eta$ (commonly measured in units of millipascal seconds or centipoise, cP = $10^{-3}$ Pa$\,\cdot\,$s $\approx 7.287 \times 10^{10}$ eV$^3$ in natural units) and the \emph{kinematic} viscosity $\nu$, defined as the ratio of $\eta$ to the liquid density (commonly measured in units of centistokes, cSt = $10^{-6}\m{m^2/s} \approx 1.691 \times 10^{-8}$ eV$^{-1}$ in natural units). The kinematic viscosity allows for the fact that certain liquids resist changes to their flow simply by being very dense. It controls the rate of diffusion of momentum in a moving liquid, and is arguably at least as important as $\eta$. (In relativistic hydrodynamics, the same remarks apply, though the relativistic versions of these quantities may behave in unusual ways, for example in response to variations of temperature.)

Which is more viscous, water or mercury? There is no conclusive answer: the respective dynamic viscosities of water and mercury \cite{kn:merc} are (at around 290 kelvin) 1.002 cP and 1.53 cP, but their respective kinematic viscosities (at that temperature) are 1.004 cSt and 0.12 cSt. Similarly, while it is often stated that ``the viscosity'' of the QGP is ``small'', this invariably refers to $\eta_s$. By everyday standards, $\eta$ itself for the QGP is far from small, being probably just under $5 \times 10^{14}$ cP $\approx 3.64 \times 10^{25}$ eV$^3$ \cite{kn:tean} for the plasmas produced in collisions at the RHIC facility, and even more for LHC plasmas. On the other hand, the \emph{kinematic} viscosity of this plasma \emph{is} small compared to that of most ``ordinary'' liquids, being (see below for the calculation) just under 0.07 cSt ---$\,$ about the same as that of mercury at 600 kelvin \cite{kn:merc}, this being one of the lowest measured values for any liquid.

It seems that these comparisons are not very helpful. In any case they are not of fundamental significance; nor are they really relevant, in view of the fact that the QGP itself only exists under conditions which are far indeed from being ``everyday'' or ``ordinary'': we are dealing with a liquid moving, at a speed equal to a substantial fraction of the speed of light ---$\,$ and one knows, as above, that relativistic hydrodynamics differs in some ways from classical hydrodynamics ---$\,$ through a ``tube'' with a radius measured in femtometres.

In hydrodynamics, the magnitude of the viscosity of a liquid is usually assessed in a different way, which we propose, following Csernai et al. \cite{kn:KelvinHelm}, to adopt here: it is done by evaluating the dimensionless \emph{Reynolds number}, defined as
\begin{equation}\label{REY}
Re \;=\;{u \delta\over \nu},
\end{equation}
where $u$ and $\delta$ are the characteristic velocity and transverse dimension of the flow, and $\nu$ is the kinematic viscosity as above. (In natural units, $u$ is dimensionless, dynamic viscosity has units of eV$^3$, energy density has units of eV$^4$, so $Re$ is indeed dimensionless since $\nu$ has the same units as $\delta$, namely eV$^{-1}$.) We can think of $Re$ as a dimensionless version of (the reciprocal of) the kinematic viscosity. It measures kinematic viscosity not by comparing it with other liquids but by including the circumstances in which the liquid in question finds itself.

As is well known, the Reynolds number has a fundamental role in hydrodynamics: most notably, it provides a remarkably useful indicator of situations in which laminar flow makes the \emph{transition to turbulence}. This happens when $Re$ is much larger than unity, typically in the thousands, and such values for $Re$ are commonly encountered even in laminar flows. That is, the kinematic viscosity is often very small relative to $u\delta$. From a physical point of view, then, one can make the case that ``small viscosity'' should generically mean that the kinematic viscosity is extremely small relative to the product of the characteristic velocity and size of the system, so that $Re$ is large, in the range of thousands or more.

Now in fact the Reynolds number of the actual QGP (at, for example, the RHIC facility) is expected to be very \emph{small} by these standards. Csernai et al. \cite{kn:KelvinHelm} estimate values in the range $3 - 10$. We shall use an upper bound corresponding to $\eta \approx 3.64 \times 10^{25}$ eV$^3$ as given in \cite{kn:tean}, which, combined with the standard \cite{kn:phobos} estimate of the energy density of the RHIC plasma at the relevant time, $\approx 4$ GeV/fm$^3 \approx 3.058 \times 10^{34}$ eV$^4$, yields
\begin{equation}\label{QGP}
\nu(QGP)\;\leq \; \approx 1.19 \times 10^{-9} \m{eV}^{- 1};
\end{equation}
this is the value 0.07 cSt mentioned above. With the estimates of $u$ and $\delta$ used in \cite{kn:KelvinHelm}, $u \approx 0.4$ and $\delta \approx 5$ fm, we have
\begin{equation}\label{REQGP}
Re(QGP)\;\geq \; \approx 8.52,
\end{equation}
which is indeed small (and compatible with the range given in \cite{kn:KelvinHelm}). Thus, in this objective sense, the QGP should be considered to be very viscous\footnote{The Reynolds number is by no means the only indicator of hydrodynamic instability, so its smallness here does not establish that the flow of the  QGP is fully stable: see \cite{kn:KelvinHelm} for a detailed phenomenological discussion of the QGP Reynolds number in this context. The $3 - 10$ range cited there does not, of course, take into account the unusual effects predicted in the present work: for the sake of clarity, we will take these values to apply to central or near-central collisions \emph{only}. We will argue that values as much as 8 times higher are possible for certain peripheral collisions. But these are still ``small'' Reynolds numbers.}.

The unusual smallness of $Re$ for the QGP, and its variation in response to changes of other physical parameters, need to be accounted for in a holographic description, no less than the small value of $\eta_s$. For $\nu$ clearly represents the viscosity of the plasma at least as well as $\eta_s$. Henceforth our focus will be on $\nu$ and $Re$ (though of course $\eta$ can be computed at any point in the discussion if the energy density is known).

This is not to say that $Re$ or $\nu$ should \emph{replace} $\eta_s$ as a quantity of basic interest in the study of the QGP; we will argue instead that it can \emph{complement} $\eta_s$ in some extremely useful ways.

In particular, the universality of (\ref{A}) is a two-edged sword: while it encourages the belief that (\ref{A}) is no mere peculiarity of some specific (and inevitably not fully realistic) model, it also means that completely \emph{unrealistic} holographic models can correctly ``predict'' $\eta_s$. In other words, a bad prediction for $\eta$ (and therefore for $\nu$) can be compensated by an equally bad prediction for the entropy density. One expects that a holographic model of $\nu$ can detect this. In this work we will be concerned precisely with just such a case; we now discuss general aspects of such a model, and the results we have obtained from it.

\addtocounter{section}{1}
\section* {\large{\textsf{2. Holographic Model of $\nu$(QGP): Generalities and Results}}}
Recently, it has become clear that any holographic model of the QGP that ignores \emph{magnetic fields} must be considered inadequate: for extremely intense magnetic fields do permeate the plasma in all but the most central collisions of heavy ions \cite{kn:skokov,kn:tuchin,kn:magnet,kn:review,kn:hattori}. The field experienced by the plasma in any given case depends on the impact parameter, $b$; in a specific beam it varies with $b$ from zero up to some maximum, which can be computed.

The magnetic field is incorporated in a holographic model simply by attaching a magnetic charge to the bulk black hole responsible for the thermal properties of the bulk theory, and of course one need not go beyond Einstein gravity to implement this. However, because the magnetic charge generates a bulk field that back-reacts on the black hole geometry, this procedure affects the position of the event horizon, thereby altering the relation between the mass of the black hole and its entropy, as well as the relation between these parameters and the Hawking temperature. Therefore, the holographically dual versions of all of these quantities will be sensitive to the magnetic field. In one extraordinary case, $\eta/s$, the effects cancel (as long as we retain Einstein gravity in the bulk), but one must not expect this to occur for other combinations such as $\nu$ and $Re$, and indeed it does not.

In short, models that ignore magnetic fields may well predict $Re$ badly, while nevertheless being fully compatible with (\ref{A}). The question is this: in concrete cases, does the holographically predicted value of $Re$ vary \emph{significantly} with the magnetic field, at a given value of the temperature? The results will be model-dependent, but this just means that we should, as always, begin with a simple model and work towards more complex ones, hoping that qualitative features of the former will persist into the latter.

Simplicity in this case means that we wish to keep $\eta_s$ fixed at $1/4\pi$, so that we have a clean indication of the effects of the magnetic field on $Re$, and because holography itself predicts (see below) that, even if it is allowed to vary, $\eta_s$ does so extremely slowly even for very large values of the field.

We can achieve this by using an explicit bulk black hole metric, which solves the Einstein equations \emph{exactly}, and which incorporates the minimal set of thermodynamic and other variables\footnote{Far more sophisticated models are available: see for example \cite{kn:dud,kn:gurs} and references therein.} needed here. We therefore exclude, in the present work, all higher-derivative corrections to the KSS bound; thus, for example, we do not allow for the theoretically indicated possible slow increase \cite{kn:sera2} of $\eta_s$ with temperatures from the lower end of the hadronic/plasma crossover up to temperatures characteristic of the QGP\footnote{See \cite{kn:braguta,kn:fei,kn:luzum} for recent (non-holographic) theoretical perspectives on this question. Results from the ALICE experiment \cite{kn:YZ} indicate however that, for the range of temperatures we shall consider here, this variation is not significant.}; for the same reason we are excluding variations of $\eta_s$ with the baryonic chemical potential \cite{kn:myers}: that is, we consider only the case $\mu_B = 0$.

Simplicity also means that we focus attention purely on viscosity \emph{in the reaction plane} of a given heavy-ion collision: in the conventional coordinates, this means that we consider only the two-dimensional $(x, z)$ plane, where $z$ corresponds to the axis of the collision, and not viscous flow in the perpendicular $y$ direction ---$\,$ the direction of the magnetic field. Apart from the usual benefits \cite{kn:horror,kn:hartlusach} of restricting the dimensionality in a holographic treatment, this is particularly important here because recent work has suggested that magnetic fields can give rise to violations of the KSS bound, if one attempts to apply it to shearing in the $y$ direction. (The point is that the KSS bound was originally derived in the context of full ($SO(3)$) symmetry, whereas the magnetic field preserves only the $SO(2)$ symmetry of the reaction plane: see \cite{kn:erd,kn:mam,kn:crit,kn:triv,kn:rogat}.) We will maintain $SO(2)$ symmetry (and translation symmetry \cite{kn:sean,kn:bag,kn:poov}) in the reaction plane here, so the KSS bound is not affected by the presence of a strong magnetic field\footnote{In any case, even in the direction of the field, the holographically predicted effect of the magnetic field on $\eta_s$ is small: see \cite{kn:crit}.}.

Thus, while in the most general case $\eta_s$ can vary with temperature, direction, and perhaps other parameters, we confine ourselves here to the simplest cases, in which it does not: so we have equation (\ref{A}) and the usual corresponding inequality at finite coupling. We will then use this to estimate or put a lower bound on the kinematic viscosity, or an upper bound on the plasma Reynolds number, by means of a holographic computation of the ratio of the entropy density to the energy density. This can be done by means of a relatively straightforward investigation of a suitable black hole geometry in the bulk. It is reasonable to expect that qualitative results obtained in this manner will persist in more complex, less tractable models.

We find that holography imposes, in the absence of a magnetic field (that is, for central or near-central collisions), a surprisingly strong upper bound on the Reynolds number of the QGP studied at the RHIC facility: using the same values for $u$ and $\delta$ as in \cite{kn:KelvinHelm} (leading to the phenomenological estimate given there, $Re \approx 3 - 10$), we find a holographic bound of the form
\begin{equation}\label{REYSTIMATE}
Re_0^{\m{fc}}(\m{Hol})\;<\;Re_0(\m{Hol})\; \approx \;20,
\end{equation}
where the subscript indicates a zero magnetic field and the superscript fc denotes the value at finite coupling\footnote{We are assuming here that the effect of finite coupling is always to increase the kinematic viscosity. The (admittedly heuristic) reasoning here is based on simple concrete examples, such as \cite{kn:yaffe} the case of the quartically coupled scalar, with coupling term $g\phi^4$. Here (see \cite{kn:nat}, Section 12.1.2) one finds that $\eta$ scales with $T^3/g^2$, and the energy density with $T^4$; so both $\eta$ and $\nu$ should be expected to be larger for finite than for infinite coupling. This argument certainly should be improved; it does not show, for example, that our bound will be as realistic as that of KSS.}. \emph{Thus, holography correctly encodes the smallness of the Reynolds number in this case;} and the holographic bound is in very good agreement with the estimated \emph{lower} bound on the Reynolds number of the actual QGP given by the inequality (\ref{REQGP}). (Bear in mind that ``typical'' Reynolds numbers for laminar flows are in the hundreds or thousands.)

Next we turn to the case of peripheral collisions, in which the magnetic field can be strong. But before discussing our results, let us first ask: what do we expect?

Qualitatively, we expect the kinematic viscosity to decline as the magnetic field increases: kinematic viscosity corresponds to a momentum transfer (in fact, $\nu$ is sometimes called the ``momentum diffusivity''), and this transfer will be suppressed by a magnetic field in directions perpendicular to the field \cite{kn:kitu}.

More interestingly, a decline is also expected in connection with the remarkable phenomenon of \emph{paramagnetic squeezing} \cite{kn:bali}. Briefly, the magnetic fields exerted on the plasma in peripheral collisions will tend to compress the plasma in the reaction plane directions (and to elongate it in the $y$ direction, the direction of the field). This directly affects the pressure gradients in the plasma and may well have very important consequences for the value of the elliptic flow parameter $v_2$. However, it can also be expected to have an effect relevant here, by reducing the dominant outward pressure in the reaction plane; and such a reduction will normally\footnote{One should be somewhat cautious here, because there are anomalous liquids for which, in certain regimes, a reduction of pressure gives rise to an \emph{increase} in viscosity. A perhaps surprising example of such a liquid is \emph{water}: for temperatures below $+32°$C, water's viscosity does indeed increase as the pressure drops from around 20 megapascals. However, this behaviour is associated (in a very complex way, see \cite{kn:chem}) with a better-known anomaly of water, its negative thermal expansion coefficient in certain regimes. One does not expect such behaviour in the QGP.} result in a reduced viscosity.

However, it is a general rule in hydrodynamics that variations of viscosity with variables other than temperature are extremely small, even for large changes in such variables. Thus, while a decrease in the kinematic viscosity of the QGP is to be expected (for either of the above reasons) for collisions involving increasing magnetic field values, it is very unclear whether this decrease will be significant.

We find that holography predicts that the kinematic viscosity of the QGP, at fixed temperature, does decline with increasing magnetic fields: so at least the direction of the predicted variation is correct. Equally important, the variation is quite large, \emph{even for relatively small\footnote{Of course, the reader should bear in mind that the ``relatively small'' magnetic fields here are measured in units of $m_{\pi}^2 \approx 10^{18}$ gauss.} values of the magnetic field}: for example, for plasmas produced at the RHIC facility, an increase of the field from zero to $eB \approx 2 \times\,m_{\pi}^2$ (where $m_{\pi}$ is the conventional pion mass) causes a change in the predicted kinematic viscosity of almost 25$\%$; for $eB \approx 4 \times\,m_{\pi}^2$, the change is nearly 50$\%$. (Of course, the corresponding Reynolds numbers increase accordingly.) Such magnetic fields are well below the latest estimates for the maximal fields produced in peripheral collisions at the RHIC, which range up to $eB \approx 10\,m_{\pi}^2$ \cite{kn:fivetimes,kn:bzdak,kn:denghuang,kn:shipu,kn:holliday}, and are therefore quite realistic, in the sense that they can be attained for a wide variety of peripheral collisions (and not merely those with impact parameter carefully selected to achieve the maximal value). The predicted variations of the kinematic viscosity and Reynolds number in the case of the plasmas observed by the ALICE \cite{kn:alice} and other experiments at the LHC are still more dramatic.

This in turn means that the role of magnetic fields in, for example, discussions of paramagnetic squeezing, may be more complex than expected. For if the magnetic field not only squeezes the plasma, but at the same time reduces its kinematic viscosity, one can expect that the flow will respond more effectually to that squeezing. It would be of interest to consider this magnetically induced amplification effect in connection with searches for observational evidence of paramagnetic squeezing \cite{kn:gergend}.

In short: holography suggests that the Reynolds number of the QGP may have a surprisingly strong dependence on the centrality of the collision in which it is produced.

In summary, then, a holographic study of the Reynolds number or kinematic viscosity $\nu$ does in fact usefully complement the more familiar holographic universality of the entropy-normalized viscosity $\eta_s$. For in a simple but very explicit holographic model, one finds an upper bound on $Re$ for central or near-central collisions which is in good agreement with the observational lower bound, and that, whereas $\eta_s$ does not vary with the collisional impact parameter (that is, with the magnetic field), $\nu$ does vary in a sense which is physically reasonable but with a surprisingly large magnitude.

We now explain the details.

\addtocounter{section}{1}
\section* {\large{\textsf{3. Holographic Model of $\nu$(QGP): Specifics}}}
For our purposes, the simplest possible bulk black hole \cite{kn:dyon} has a flat (indicated by a zero superscript) planar event horizon and a Euclidean asymptotically AdS magnetic Reissner-Nordstr\"om metric $g^E(\m{AdSP^*RN^{0})}$, given by\footnote{In the bulk, we work exclusively in the Euclidean domain. The Lorentzian case involves, in general, additional subtleties, not relevant here (because magnetic charge is not complexified when passing to the Lorentzian domain): see \cite{kn:84} and the discussion below.}
\begin{eqnarray}\label{B}
g^E(\m{AdSP^*RN^{0})} & = &  \Bigg[{r^2\over L^2}\;-\;{8\pi M^*\over r}+{4\pi P^{*2}\over r^2}\Bigg]\m{d}t^2\; \nonumber \\
& &  + \;{\m{d}r^2\over {\dfrac{r^2}{L^2}}\;-\;{\dfrac{8\pi M^*}{r}}+{\dfrac{4\pi P^{*2}}{r^2}}} \;+\;r^2\left[\m{d}\psi^2\;+\;\m{d}\zeta^2\right];
\end{eqnarray}
here $L$ is the asymptotic AdS curvature scale, $r$ and $t$ are the usual radial and ``time'' coordinates, and $M^*$ and $P^*$ are parameters such that, if $r_h$ is the value of the radial coordinate at the event horizon, then, if $\ell_P$ is the bulk Planck length, $M^*/\ell_P^2r_h^2$ is the mass per unit horizon area, and $P^*/\ell_Pr_h^2$ is the magnetic charge per unit horizon area; and $\psi$ and $\zeta$ are dimensionless coordinates on the plane, related at infinity to the standard coordinates $x = L\psi$ and $z=L\zeta$ in the reaction plane (the $y$ axis being parallel to the magnetic field) of a heavy-ion collision\footnote{To understand the significance of this, note that we are entitled to regard $\psi$ and $\zeta$ as angular coordinates on a flat torus (though this is not essential, and it plays no further role here). Clearly, $L$ sets the length scale of the boundary theory once conformal symmetry is broken. We therefore take it to have a typical nuclear physics value, 10 fm.}. The Hawking temperature of this black hole defines the temperature $T_{\infty}$ of the boundary theory, and the parameter $P^*$ determines the boundary\footnote{We will use $B_{\infty}$ to refer to the ``holographic'' magnetic field, the field on the boundary of the relevant asymptotically AdS spacetime. Observed or phenomenologically computed magnetic fields will be denoted simply by $B$.} magnetic field $B_{\infty}$ \cite{kn:hartkov}, in the familiar ways.

We note that the bulk being four-dimensional, the boundary theory has two-dimensional spatial sections: these are taken to represent the reaction plane. That is, we are constructing a holographic model of the relation between magnetic fields and viscosity in that plane \emph{only}. Precisely because the magnetic fields we consider are so intense, and because the relative motion between different sections of the form $y = \m{constant}$ is relatively negligible, and also (see above) because we take the physics in the reaction plane to be isotropic, we assume that this is a good approximation; note however that in other cases (such as recently proposed experiments \cite{kn:ZRRU} involving collisions of highly non-spherical nuclei), it might not be.

Unfortunately, this model is in fact a little \emph{too} simple. To see why, recall that this bulk geometry corresponds to a simplified version of a much more complex string-theoretic configuration. This simplification is legitimate \emph{provided} that it does not cause the more complex system to become internally inconsistent. The relevant consistency conditions have been discussed in \cite{kn:ferrari1,kn:ferrari2,kn:ferrari3,kn:ferrari4}. One of these conditions (discussed in detail in \cite{kn:ferrari3}) is that a certain function $\mathfrak{S^{\m{E}}}$ defined on the bulk geometry (in terms of the areas and volumes of hypersurfaces homologous to the conformal boundary) should never be negative at any point\footnote{The Lorentzian version of the argument runs as follows \cite{kn:84}: the Lorentzian version of $\mathfrak{S^{\m{E}}}$ is, up to factors involving the brane tension, equal to the action of a BPS brane in the Lorentzian black hole geometry \cite{kn:seiberg}. If this action becomes negative at sufficiently large $r$, this means that it is smaller than its value at the event horizon. The black hole will therefore generate arbitrary quantities of branes by a pair-production process \cite{kn:maldacena, kn:KPR} at such values of $r$, and these branes will have no tendency to contract back into the hole. The system therefore becomes unstable.}. In \cite{kn:82} we translated this inequality into an inequality governing the parameters of the specific metric $g^E(\m{AdSP^*RN^{0})}$, and we found that this condition (when the baryonic chemical potential vanishes, as we assume throughout) takes a very simple form:
\begin{equation}\label{C}
B_{\infty}\;\leq \;2\pi^{3/2}T_{\infty}^2\;\approx \; 11.14 \times T_{\infty}^2.
\end{equation}
That is, if this inequality is violated, then $\mathfrak{S^{\m{E}}}$ evaluated on this geometry does take on negative values and there is a serious mathematical inconsistency.

This inequality is in fact satisfied by the plasmas produced in most heavy-ion collisions at the RHIC facility, but (in view of recent suggestions \cite{kn:fivetimes,kn:bzdak,kn:denghuang,kn:shipu,kn:holliday} that the magnetic fields produced in these collisions may be even larger than previously thought) probably not in all; and it is still more likely to be violated in collisions at the LHC (again, particularly with the recent upward revision in estimates of $B$). For example, as mentioned above, the latest estimates for the maximal magnetic field produced in RHIC collisions put it around $eB \approx 10 \times m_{\pi}^2$, or $B = 6.46 \times 10^5$ MeV$^2$; whereas for typical RHIC temperatures ($T \approx 220$ MeV), the right side of (\ref{C}) is $\approx 5.42 \times 10^5$ MeV$^2$.

In \cite{kn:88} we argued that the problem here is that the computations in \cite{kn:82} (which was concerned with a different application, to cosmological magnetic fields) neglected an important physical effect: the fact that these collisions involve not just large magnetic fields, but also large angular momentum densities, as predicted in \cite{kn:liang,kn:bec,kn:huang,kn:viscous,kn:deng,kn:vortical}, and possibly observed by the STAR collaboration \cite{kn:STARcoll}. It turns out that the inclusion of an angular momentum parameter in the bulk geometry can resolve the consistency problem if it is sufficiently large\footnote{There is an issue here: while consistency can always be re-established in this way, it is not clear whether the required value of the angular momentum parameter is realistic in the case of collisions with maximal magnetic fields. We found in \cite{kn:88} (see also \cite{kn:89}) that this is definitely not a problem in the case of the RHIC plasmas, but that it could be in the LHC case. Since the phenomenon in which we are interested here appears already even at relatively low values of the magnetic field, we will not take a stand on this issue here.}.

Unfortunately, the bulk geometry used in \cite{kn:88} was extremely intricate; the metric \cite{kn:shear} generalizes the (already very complicated) topological black hole metric of Klemm et al. \cite{kn:klemm}. (It is a member of the very general Pleba\'nski--Demia\'nski family of metrics \cite{kn:plebdem,kn:grifpod}.) More importantly, it is anisotropic in the reaction plane. This anisotropy may well ultimately have to be taken into account, but it would complicate the picture still further; recall that we wish to maintain the residual SO(2) symmetry in the reaction plane.

A far simpler and more tractable way to restore the mathematical consistency of the system is to introduce a \emph{dilaton} into the bulk. This is a very natural move from a string-theoretic point of view; it need not break the SO(2) symmetry in the reaction plane, and it can be done, with a careful choice of potential, in such a way that the geometry remains asymptotically AdS \cite{kn:gz}: see \cite{kn:gout} for more detail, including the embedding in a higher-dimensional theory.

With a sufficiently strong coupling to the magnetic field\footnote{The coupling takes the form $e^{-2\alpha \varphi}F^2$, where $F$ is the field strength two-form, $\varphi$ is the dilaton, and $\alpha$ is a coupling constant.}, the dilaton has much the same effect as angular momentum \cite{kn:ong}, that is, it enforces the consistency condition of \cite{kn:ferrari3} even when the inequality (\ref{C}) is violated. We can think of the dilaton coupling constant $\alpha$ as a sort of proxy for the angular momentum parameter (at least for impact parameters which are not very large).

To see how this works in detail, consider the metric of an asymptotically AdS planar dilatonic Euclidean Reissner-Nordstr\"om black hole with magnetic charge parameter $P^*$ and mass parameter $M^*$: it takes the form \cite{kn:gz}
\begin{equation}\label{D}
g^E(\m{AdSdilP^*RN}^{0})=U(r)\m{d}t^2 + {\m{d}r^2\over U(r)} + [f(r)]^2 \left[\m{d}\psi^2\;+\;\m{d}\zeta^2\right],
\end{equation}
where the coordinates are as above; here
\begin{equation}\label{E}
U(r)=-\frac{8\pi M^*}{r}\left[1-\frac{(1+\alpha^2)P^{*2}}{2M^*r}\right]^{\frac{1-\alpha^2}{1+\alpha^2}} + \frac{r^2}{L^2} \left[1-\frac{(1+\alpha^2)P^{*2}}{2M^*r}\right]^{\frac{2\alpha^2}{1+\alpha^2}},
\end{equation}
and
\begin{equation}\label{F}
f(r)^2 = r^2\left(1-\frac{(1+\alpha^2)P^{*2}}{2M^*r}\right)^{\frac{2\alpha^2}{1+\alpha^2}}.
\end{equation}
As before, $M^*$ and $P^*$ are related to the mass and magnetic charge per unit event horizon area: if $r = r_h$ at the event horizon, then the mass per unit event horizon area is $M^*/(\ell_P^2f(r_h)^2),$ while $P^*/(\ell_Pf(r_h)^2)$ is the magnetic charge per unit event horizon area.

The dilaton itself, $\varphi$, is described by a potential of the form
\begin{equation}\label{G}
V(\varphi)\;=\;{- 1\over 8\pi \ell_P^2L^2}{1\over (1+\alpha^2)^2}\left[\alpha^2\left(3\alpha^2-1\right)e^{-2\varphi/\alpha}\,+\,\left(3-\alpha^2\right)
e^{2\alpha \varphi}\,+\,8\alpha^2e^{\alpha \varphi - \left(\varphi/\alpha\right)}\right],
\end{equation}
carefully constructed so as to preserve the asymptotic AdS character of $g^E(\m{AdSP^*RN^{0})}$; note that, since the dilaton itself is given here by
\begin{equation}\label{H}
e^{2\alpha \varphi (r)}\;=\;\left(1-\frac{(1+\alpha^2)P^{*2}}{2M^*r}\right)^{\frac{2\alpha^2}{1+\alpha^2}},
\end{equation}
we see that $V(\varphi) \rightarrow -\, 3/\left(8\pi \ell_P^2L^2\right)$, the AdS energy density, when either $\alpha \rightarrow 0$ or $r \rightarrow \infty$. That is, the asymptotic AdS energy density is embedded in $V(\varphi)$. (The kinetic term for the dilaton is the standard one.)

The electromagnetic field two-form in this case takes the form
\begin{equation}\label{I}
F\;=\;{P^*\over \ell_P}\left(1-\frac{(1+\alpha^2)P^{*2}}{2M^*r}\right)^{\frac{2\alpha^2}{1+\alpha^2}}\,\m{d}\psi \wedge \m{d}\zeta.
\end{equation}
This can be used to define a magnetic field at infinity given by
\begin{equation}\label{J}
B_{\infty}\;=\; {P^*\over L^3}.
\end{equation}

The function $U(r)$ vanishes at $r = r_h$, which is therefore to be found by solving
\begin{equation}\label{K}
-\frac{8\pi M^*}{r_h}\left[1-\frac{(1+\alpha^2)P^{*2}}{2M^*r_h}\right]^{\frac{1-\alpha^2}{1+\alpha^2}} + \frac{r_h^2}{L^2} \left[1-\frac{(1+\alpha^2)P^{*2}}{2M^*r_h}\right]^{\frac{2\alpha^2}{1+\alpha^2}} = 0,
\end{equation}
and the Hawking temperature, which determines the temperature at infinity, is
\begin{eqnarray}\label{L}
4\pi T_{\infty}&=&{8\pi M^*\over r_h^2}\left(1-\frac{(1+\alpha^2)P^{*2}}{2M^*r_h}\right)^{{1-\alpha^2\over 1+\alpha^2}}\;-\;{4\pi (1-\alpha^2)P^{*2}\over r_h^3}\left(1-\frac{(1+\alpha^2)P^{*2}}{2M^*r_h}\right)^{{-2\alpha^2\over 1+\alpha^2}}\;\nonumber \\ &
&+\;{2r_h\over L^2}\left(1-\frac{(1+\alpha^2)P^{*2}}{2M^*r_h}\right)^{{2\alpha^2\over 1+\alpha^2}}\;+\;{\alpha^2P^{*2}\over M^*L^2}\left(1-\frac{(1+\alpha^2)P^{*2}}{2M^*r_h}\right)^{{\alpha^2 - 1 \over 1+\alpha^2}}.
\end{eqnarray}

Finally, for this bulk geometry, the function $\mathfrak{S^{\m{E}}}$ discussed earlier is given \cite{kn:ong}\cite{kn:89}, up to a positive constant factor, by
\begin{eqnarray}\label{M}
\mathfrak{S^{\m{E}}}(\m{AdSdilP^*RN^{0}})(r) & = &\frac{r^3}{L}\left[1-\frac{(1+\alpha^2)P^{*2}}{2M^*r}\right]^{\frac{3\alpha^2}{1+\alpha^2}}
\left[1-\frac{8\pi M^*L^2}{r^3}\left(1-\frac{(1+\alpha^2)P^{*2}}{2M^*r}\right)^{\frac{1-3\alpha^2}{1+\alpha^2}}\right]^{\frac{1}{2}}\nonumber \\ & &-\frac{3}{L}\int_{r_h}^r s^2\left[1-\frac{(1+\alpha^2)P^{*2}}{2M^*s}\right]^{\frac{2\alpha^2}{1+\alpha^2}} \m{d}s.
\end{eqnarray}

Given values of $B_{\infty}$, $T_{\infty}$, $L$, and $\alpha$, one can solve\footnote{In fact, if $\alpha$ is chosen too large, this system of simultaneous equations turns out to have no real solutions. This interesting phenomenon will not be an issue in this work, where we always choose $\alpha$ to be as small as possible consistent with the non-negativity of $\mathfrak{S^{\m{E}}}$.} the three equations (\ref{J}), (\ref{K}), and (\ref{L}) for $r_h$, $P^*$, and $M^*$, and then equation (\ref{M}) specifies $\mathfrak{S^{\m{E}}}(\m{AdSdilP^*RN^{0}})(r)$ completely; one can then check directly (by simply graphing it) whether this function is strictly non-negative as $r$ increases from $r_h$ to infinity.

We found in \cite{kn:88} that, for the ``maximal'' RHIC data used above ($eB = 10 \times\,m_{\pi}^2$, $T \approx 220$ MeV), then $\mathfrak{S^{\m{E}}}$ for the (much more complex) geometry used there is forced to be positive everywhere when even a very small amount of shearing angular momentum is added to the bulk black hole. In the same way, with these values for $B_{\infty}$ and $T_{\infty}$, the positivity of $\mathfrak{S^{\m{E}}}$ for the dilatonic black hole geometry can be restored with a dilaton that is very weakly coupled\footnote{See \cite{kn:89} for the precise meaning of ``weak coupling''. Essentially it means that $\alpha$ is much smaller than its maximal possible value (for these data), which is around 0.720. See the preceding footnote.}: one only needs to take $\alpha = 0.06$, as can be seen in Figure 1.

\begin{figure}[!h]
\centering
\includegraphics[width=0.65\textwidth]{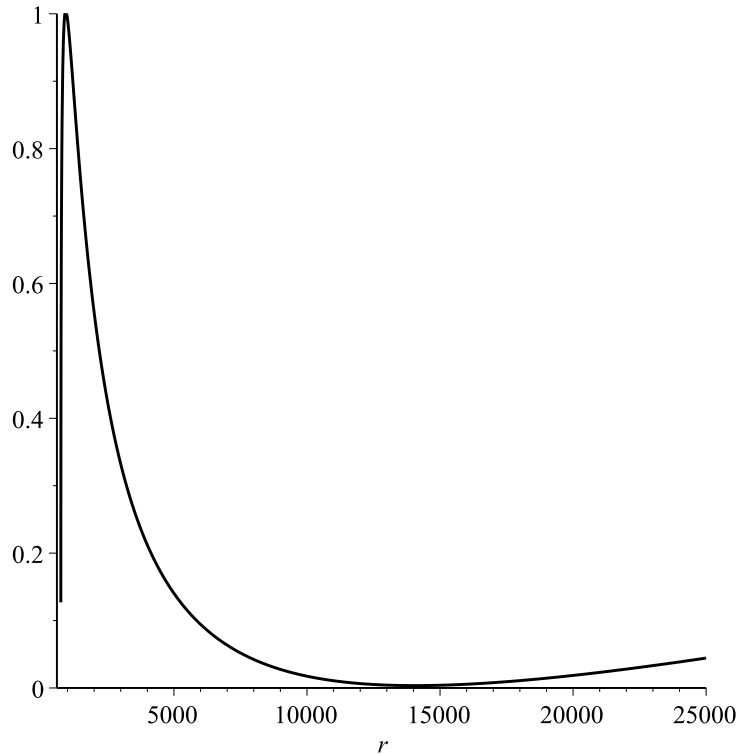}
\caption{$\mathfrak{S^{\m{E}}}(\m{AdSdilP^*RN^{0}})(r)$, $T_{\infty} \approx 220$ MeV, $\mu_B = 0$, $eB_{\infty} \approx 10 \times m_{\pi}^2$, $\alpha = 0.06$.}
\end{figure}

On the other hand, for maximal LHC data, far larger values of the angular momentum were required to ensure that $\mathfrak{S^{\m{E}}}$ is never negative in that case; similarly, we found in \cite{kn:89} that, for these data, a dilaton coupling at least as large as $\alpha \approx 0.284$ is needed. Thus, the dilaton coupling does mimic the effect of angular momentum for our purposes\footnote{The precise reason for the fact that the dilaton tends to restore consistency in this way is not completely known. As mentioned earlier, the bulk interpretation of the inconsistency involves uncontrolled pair-production of branes. The location of a brane in the bulk can be represented \cite{kn:seiberg} by a scalar field at infinity, which (because of the form of the conformal coupling term) behaves well if the scalar curvature at infinity is positive, but it misbehaves if the scalar curvature at infinity is negative, and \emph{sometimes} if it vanishes, as it does here. The dilaton turns on a source and a vacuum expectation value for the dual operator (in fact as a multi-trace operator, as in \cite{kn:multiwit}). In some way that remains to be understood in detail, this has the effect (if the coupling is strong enough) of pushing the scalar field to behave more like the positive case. Understanding this in detail would require a delicate analysis of high-order terms in the action given in \cite{kn:seiberg}.}.

We are now in a position to give a self-consistent holographic computation of the kinematic viscosities of the plasmas produced, in collisions of varying centrality, at the RHIC and LHC facilities.

\addtocounter{section}{1}
\section* {\large{\textsf{4. The Kinematic Viscosity as a Function of the Magnetic Field}}}
We now wish to compute the kinematic viscosity, using the assumed holographic duality of the above AdS planar dilatonic Reissner-Nordstr\"om black hole with a field theory that models the QGP when it is subjected to a non-zero magnetic field.

We saw above that the mass (that is, the energy) of the black hole per unit event horizon area is given by $M^*/(\ell_P^2f(r_h)^2),$ where $f(r)$ is given by equation (\ref{F}). The entropy per unit horizon area is (for Einstein gravity, as assumed here, in natural units) $1/4\ell_P^2$. Assuming uniformity in the $y$ direction on the boundary, the ratio of the field theory entropy density to its energy density is then given holographically by
\begin{equation}\label{N}
{s\over \rho}\;=\;{f(r_h)^2\over 4M^*};
\end{equation}
notice that the bulk quantity on the right has no dependence on $\ell_P$. Combining this with equations (\ref{A}) and (\ref{F}), we have finally, since $\nu = \eta/\rho$,
\begin{equation}\label{O}
\nu \;=\; {r_h^2\left(1-\frac{(1+\alpha^2)P^{*2}}{2M^*r_h}\right)^{\frac{2\alpha^2}{1+\alpha^2}}\over 16\pi M^*}.
\end{equation}

The right hand side of this relation involves only bulk parameters; but, as we saw above, holography (in the form of the equations (\ref{J}), (\ref{K}), and (\ref{L})) allows us to regard $r_h$, $P^*$, and $M^*$ as nothing but (very complicated) functions of the known (that is, fixed by considerations of boundary physics) parameters $B_{\infty}$, $T_{\infty}$, $L$, and $\alpha$. (As usual, equality holds here strictly only at arbitrarily large coupling: in general, one should interpret the right side of (\ref{O}) as a lower bound on the kinematic viscosity, implying an upper bound on the Reynolds number.)

Clearly, neither $\nu$ nor its dimensionless version $Re$ enjoy the universality of the holographic prediction for $\eta_s$; but we are now, with equation (\ref{O}), at least in a position to compute how they vary with the physical boundary parameters.

Our objective here is to study the effect of the magnetic fields on the kinematic viscosity. To that end, we first compute the quantity $\nu_0$, which is the kinematic viscosity that one expects for the plasma if magnetic fields are neglected. (Equivalently, this is the holographically predicted value for $\nu$ in the case of central or near-central collisions.) We then compute $\nu$ itself, including the effect of various non-zero values for the magnetic field $B_{\infty}$. The ratio $\nu/\nu_0$ then quantifies the effect we wish to study.

Before we proceed, we should briefly discuss the phenomenological model usually employed to estimate the magnetic field $B$ arising in peripheral heavy ion collisions, as a function of the impact parameter $b$. In \cite{kn:denghuang} it is found, using the HIJING Monte Carlo event generator (subsequently updated to HIJING++ \cite{kn:hijingupdate}), that $B$ rises, almost linearly, from zero (for exactly central collisions) up to a maximum value (which is determined by the collision energy: it is much larger at the LHC than at the RHIC). This maximum occurs at a value of $b$ which is so large that one can question whether a plasma is actually formed, to any significant degree, beyond that point. As it is known that the temperature of the plasma varies much more slowly with the collision energy than the magnetic field \cite{kn:denghuang}, it is a reasonable approximation, for moderate impact parameter collisions, to assume that the temperature is independent of the impact parameter: this is how we will analyse how the kinematic viscosity varies with the magnetic field in our holographic model. This allows us to separate the effects of magnetic fields from those of varying temperatures, for example when we move up from RHIC collision energies to LHC energies. Once we have understood these effects, we will be in a position to discuss collisions with large impact parameters.

Let us begin with the RHIC data. Taking $T_{\infty} \approx 220$ MeV in this case, we allow $eB_{\infty}$ to range in steps from zero up to around 10$m_{\pi}^2$ \cite{kn:fivetimes,kn:bzdak,kn:denghuang,kn:shipu,kn:holliday}. For each choice of $eB_{\infty}$, we have chosen a value of $\alpha$ close to the minimal value capable of ensuring that $\mathfrak{S^{\m{E}}}$, evaluated using equation (\ref{M}), remains non-negative everywhere on its domain. (In fact, we find that our results vary slowly with $\alpha$, so the precise choice is not important.)

As explained above, we denote the value of $\nu$ (at the stated temperature) for $B_{\infty} = \alpha = 0$ by $\nu_0$. Using equations (\ref{J}), (\ref{K}), and (\ref{L}), one finds that in fact $\nu_0 = 3/(8\,\pi T_{\infty})$, so that in this case the predicted kinematic viscosity of the RHIC plasma satisfies (the notation fc again corresponding to the finite coupling case)
\begin{equation}\label{P}
\nu_0^{\m{fc}}\;>\;\nu_0 \;\approx \; 5.410 \times 10^{-10} \m{eV^{-1}}.
\end{equation}
With the approximate values for $u$ and $\delta$ used in \cite{kn:KelvinHelm} ($u \approx 0.4, \delta \approx 5$ fm), this yields a bound for the Reynolds number given by $Re_0^{\m{fc}} < Re_0\approx 18.77$, or let us say $Re_0^{\m{fc}} < \approx 20$. This is in very good agreement with the experimentally determined lower bound on $Re(QGP)$ for the RHIC plasmas, around 8.5 (the inequality (\ref{REQGP}) given above).

\emph{In short, a combination of experimental and holographic techniques puts the Reynolds number of the RHIC plasmas somewhere between 8.5 and 20, for near-central collisions.}

For peripheral collisions, we allow $B_{\infty}$ to be non-zero. For sufficiently \emph{small} values of the field, it is consistent to take $\alpha$ to be negligible, and in this case equations (\ref{K}), (\ref{L}), and (\ref{O}) simplify to such an extent that an analytic discussion becomes possible. In fact, a straightforward manipulation shows that one has, in this case,
\begin{equation}\label{POSSUM}
{3r_h^4\over L^2}\,-\,4\pi T_{\infty}r_h^3\,-\,4\pi B_{\infty}^2L^6\;=\;0
\end{equation}
and
\begin{equation}\label{POGUE}
{1\over 2\nu}={4r_h\over L^2}\,-\,4\pi T_{\infty}.
\end{equation}

Notice that $r_h$ is obtained by solving a quartic, with $B_{\infty}$ in the (negative) constant term. If we fix the other parameters and increase $B_{\infty}$, then, the quartic drops lower and the positive root, that is, $r_h$, shifts to the right. From equation (\ref{POGUE}) we see at once that the effect of this is to cause $\nu$ to decrease. Thus we see that, at least for relatively small magnetic fields, an increase in the field always causes $\nu$ to decrease, all else being equal. Evidently this is a basic prediction of the holographic model.

For larger magnetic fields the pattern turns out to be the same, though in this case $\alpha$ can no longer be neglected and a numerical analysis of equations (\ref{K}), (\ref{L}), and (\ref{O}) becomes necessary. We state the results for $\nu$ by comparing it with $\nu_0$ (and for $Re$ by comparing it with $Re_0$; note that this ratio does not depend on $u$ or $\delta$). The results are shown in the table and (for $\nu/\nu_0$) in Figure 2.
\begin{center}
\begin{tabular}{|c|c|c|c|}
\hline
$eB_{\infty}/m_{\pi}^2$   & $\alpha$ & $\nu /\nu_0$ & $Re/Re_0$ \\
\hline
2 & 0.012 & 0.7652  & 1.307\\
4 & 0.024 & 0.5387  & 1.856\\
6 & 0.036 & 0.4165 & 2.401\\
8 & 0.048 & 0.3442 & 2.905\\
10 & 0.06 & 0.2966 & 3.372\\
\hline
\end{tabular}
\end{center}
\begin{figure}[!h]
\centering
\includegraphics[width=0.65\textwidth]{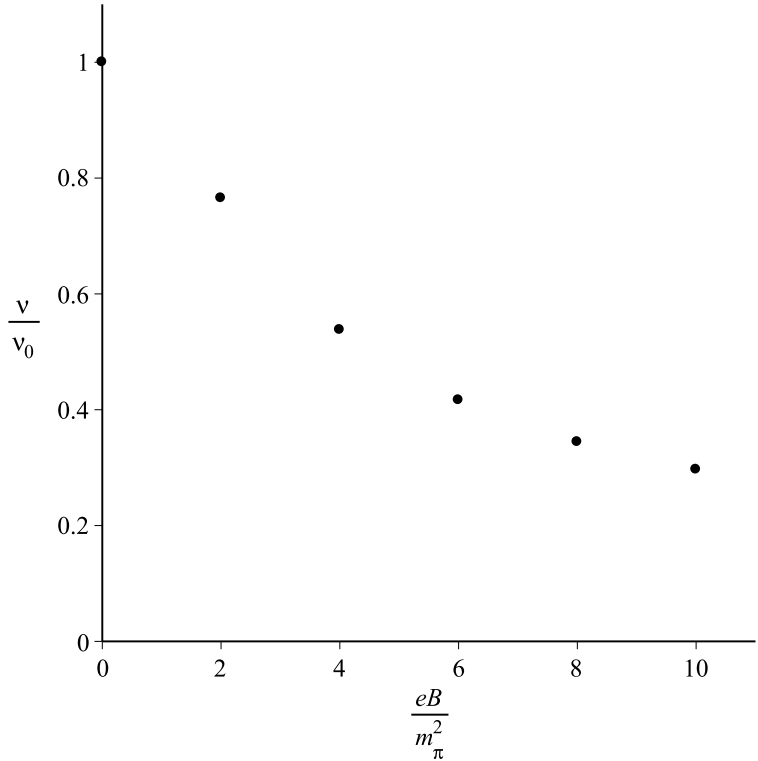}
\caption{Relative Kinematic Viscosity vs. Magnetic Field, RHIC Data.}
\end{figure}

We see that holography predicts a very substantial reduction in the kinematic viscosity when magnetic fields are taken into account; furthermore, this is true even for relatively modest values of $B_{\infty}$. Thus, for example, if we take $eB_{\infty} = 4m_{\pi}^2$, we find that $\nu$ drops to about half of its value when the magnetic field is neglected, so the Reynolds number roughly doubles\footnote{Strictly speaking, this is not the case, since we are dealing with inequalities when finite coupling is taken into account: that is, we now have $Re^{\m{fc}} < \approx 40$ instead of $Re_0^{\m{fc}} < \approx 20$. But it is natural to interpret this as jump in the predicted value of $Re^{\m{fc}}$.}. To put it another way: \emph{holography suggests that the flow of the plasma is not best described by a single Reynolds number, but rather by a range of values which depend on the centrality of the corresponding collision}. In the case of RHIC collisions, perhaps the best way to state the case is that there is an upper bound on $Re^{\m{fc}}$, ranging from about 20 for very central collisions, up to around 60 for a small subset of collisions (those at high impact parameter such that a plasma is nevertheless actually formed).

Next we turn to the LHC plasmas. Taking $T_{\infty} \approx 300$ MeV in this case, we allow $eB_{\infty}$ to range from zero up to around 70$m_{\pi}^2$ \cite{kn:denghuang,kn:gergend}. With the same notation as above, we find in this case that
\begin{equation}\label{Q}
\nu_0^{\m{fc}}\;>\;\nu_0 \;\approx \; 3.959 \times 10^{-10} \m{eV^{-1}};
\end{equation}
with the same value of $u\delta$ as before, this yields $Re_0^{\m{fc}} < \approx 26$. Thus, holography leads us to expect larger values of the Reynolds number for central or near-central LHC collisions than for their RHIC counterparts (though the value is still ``small'' in the sense we have discussed)\footnote{If the kinematic viscosity of the LHC plasmas approaches $3.959 \times 10^{-10} \m{eV^{-1}}$, which is around 0.023 cSt, this would be among the lowest kinematic viscosities of all liquids (possibly rivalled only by very dense metals with very high boiling points, such as rhenium, near their critical points).}

The results for non-zero magnetic fields are shown in the table and in Figure 3.
\begin{center}
\begin{tabular}{|c|c|c|c|}
\hline
$eB_{\infty}/m_{\pi}^2$   & $\alpha$ & $\nu /\nu_0$ & $Re/Re_0$ \\
\hline
10 & 0.04 & 0.4446 & 2.249\\
20 & 0.15 & 0.2800 & 3.571\\
30 & 0.18 & 0.2137 & 4.679\\
40 & 0.24 & 0.1758 & 5.688\\
50 & 0.27 & 0.1513 & 6.609\\
60 & 0.28 & 0.1344 & 7.440\\
70 & 0.29 & 0.1220 & 8.197\\
\hline
\end{tabular}
\end{center}
\begin{figure}[!h]
\centering
\includegraphics[width=0.65\textwidth]{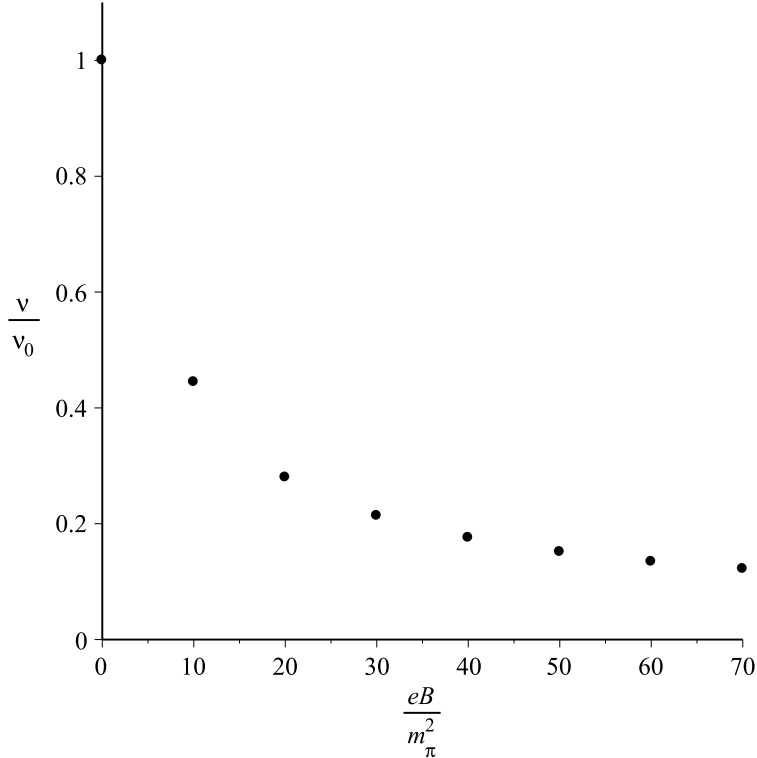}
\caption{Relative Kinematic Viscosity vs. Magnetic Field, LHC Data.}
\end{figure}

As in the RHIC case, the effect of including the magnetic field is to reduce the relative value of the kinematic viscosity, that is, to increase the Reynolds number, which again becomes centrality-dependent: the upper bound is around 26 for central or near-central collisions, but possibly up to nearly 200 for ``special'' collisions giving rise to a plasma subjected to exceptionally large magnetic fields.

Some general comments are in order here.

$\bullet$ In this work, we have focused on the kinematic viscosity, because of its relation with the most important dimensionless quantity in hydrodynamics, the Reynolds number. If one is interested in the dynamic viscosity, it can of course be reconstructed from $\nu$ simply by multiplying by the energy density. For example, a plasma resulting from a near-central collision at the RHIC facility typically \cite{kn:phobos} has an energy density, at the relevant time\footnote{The definition of the energy density in heavy ion collisions is a subtle matter: see the discussion of this issue in \cite{kn:phobos}.}, of around $\approx 4$ GeV/fm$^3 \approx 3.058 \times 10^{34}$ eV$^4$, so with the above lower bound on $\nu_0 \;\approx \; 5.410 \times 10^{-10} \m{eV^{-1}}$ we find that our model predicts a \emph{lower} bound of around $1.65 \times 10^{25}$ eV$^3$ for $\eta_0$ in this case; this is, in fact, below, but not very far below, the \emph{upper} bound on the dynamic viscosity of the actual RHIC plasma given in \cite{kn:tean}, $\approx 3.64 \times 10^{25}$ eV$^3$. For the LHC, where the lower bound on $\nu_0$ is $\approx 3.959 \times 10^{-10} \m{eV^{-1}}$ but the energy density is around 2.3 times larger \cite{kn:aliceenergy}, the predicted lower bound on $\eta_0$ is \emph{higher}, around $2.78 \times 10^{25}$ eV$^3$. Notice that the LHC plasmas are predicted to have a lower kinematic viscosity, but a higher dynamic viscosity, than their RHIC counterparts: this behaviour with increased temperature is to be expected for dense relativistic liquids.

$\bullet$ The surprisingly good agreement of $Re_0$ with observational/phenomenological expectations ---$\,$ that is, the prediction that it should be ``small'' ---$\,$ is partly, but only partly, due to the success of the KSS bound: for we have used (\ref{A}) to derive (\ref{O}). However, as we have stressed, there is a ``non-universal'' contribution to the right side of equation (\ref{O}), and this contribution might well have deviated strongly from order unity: indeed, we know that it does so in the presence of a strong magnetic field. In other words, equation (\ref{N}) is also necessary for our prediction of the Reynolds number to be satisfactory. (Note that, for example, $M^*$ is typically of order hundreds of thousands to millions (of fm) with these data; so it is not clear that the factor on the right side of (\ref{N}) will be of order unity.) To put it yet another way: holography produces good results not only for the entropy-normalized dynamic viscosity, but also for the kinematic and dynamic viscosities themselves.

$\bullet$ At fixed temperature, the relative kinematic viscosity (that is, $\nu/\nu_0$) always decreases with increasing magnetic field. At fixed magnetic field, it always increases with temperature.

$\bullet$ In order to avoid confusion, we have (in both the RHIC and LHC cases) kept the temperature fixed as the magnetic field is increased. As explained earlier, this is a reasonable approximation for collisions with moderate impact parameters, but it is not accurate for highly peripheral collisions. For these, the temperature is lower than in the nearly central case; unfortunately, the numerical extent of this effect is not well understood. However, as stated above, the relative kinematic viscosity decreases as the temperature is lowered (for a fixed magnetic field), so $\nu/\nu_0$ is \emph{smaller} in the peripheral case, for a given value of the magnetic field. That is, taking this effect into account will only accentuate the tendency of the magnetic field to reduce $\nu/\nu_0$. (In fact, however, the effect of this is rather small.)

$\bullet$ Taking these observations into account, one can arrive at a rough idea as to how $\nu/\nu_0$ varies as a function of the impact parameter. Let us focus on the LHC case, and recall that the range $0 - 70\, m_{\pi}^2$ on the horizontal axis in Figure 3 corresponds to a range \cite{kn:denghuang} of impact parameters $b \approx 0 - 14$ fm. We then proceed as in the construction of Figure 3, specifying lower temperatures associated with each step up in the magnetic field. In doing so, we are guided by remarks in \cite{kn:quench,kn:trouble}, to the effect that temperatures in peripheral LHC collisions approximate those in their central RHIC counterparts. We assume, for the sake of definiteness, that this begins to happen at around $b = 10$ fm. We then (somewhat arbitrarily) interpolate linearly. The results are shown in the table ($\alpha$ being chosen in the same manner as before) and in Figure 4; they are ``schematic'' in the sense that they are based on simplified assumptions regarding the way the temperature varies with the impact parameter.
\begin{center}
\begin{tabular}{|c|c|c|c|}
\hline
$b$ (fm)   & $T_{\infty}$ (MeV) & $\nu /\nu_0$ & $Re/Re_0$ \\
\hline
2 & 300 & 0.4446 & 2.249\\
4 & 280 & 0.2738 & 3.652\\
6 & 260 & 0.2005 & 4.988\\
8 & 240 & 0.1628 & 6.143\\
10 & 220 & 0.1309 & 7.639\\
12 & 200 & 0.1141 & 8.764\\
14 & 180 & 0.1015 & 9.852\\
\hline
\end{tabular}
\end{center}
\begin{figure}[!h]
\centering
\includegraphics[width=0.65\textwidth]{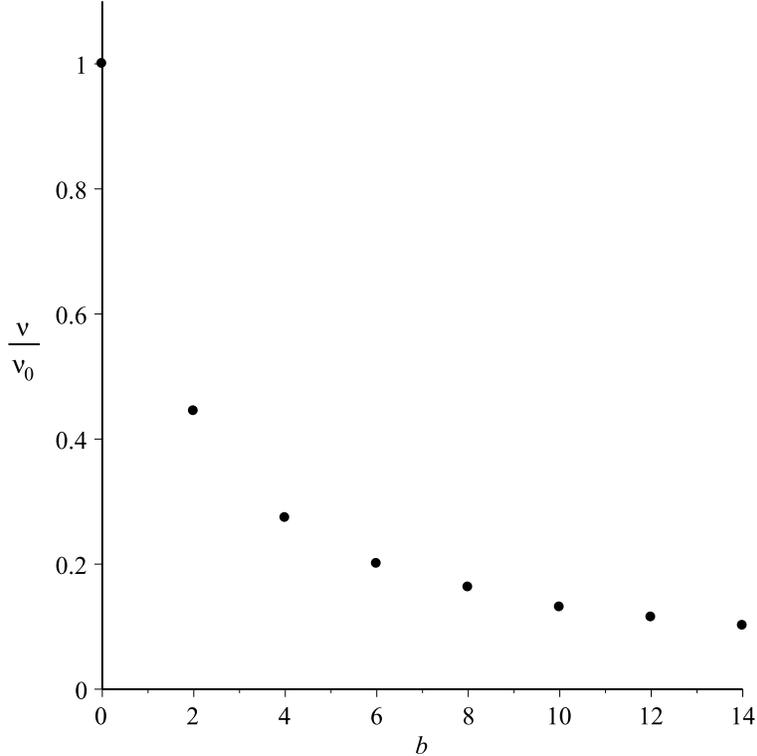}
\caption{Relative Kinematic Viscosity vs. Impact Parameter (fm), LHC Data; Schematic}
\end{figure}

The holographic model predicts that the relative kinematic viscosity drops precipitously from $b = 0$ fm to around $b = 3$ fm, but then declines much more slowly for larger impact parameters.

\addtocounter{section}{1}
\section* {\large{\textsf{5. Conclusion}}}
Much of the work on the holographic description of the QGP has (rightly) focused on the entropy-normalized dynamic viscosity, to the point where $\eta_s$ is often described as ``the viscosity'' of the QGP. Here we have argued that other measures of viscosity, notably the kinematic viscosity (or the Reynolds number) may usefully supplement the role of $\eta_s$, by being capable of responding to strong variations in physical conditions from collision to collision: in particular, $\nu$ and $Re$ can and do respond to the powerful magnetic fields which arise in peripheral collisions.

We found in fact that a very simple holographic model imposes an upper bound on $Re_0$, the Reynolds number for the QGP produced by a nearly central collision, which is in remarkably good agreement with the lower bound deduced from observations. On the other hand, the same model predicts that $Re/Re_0$, where $Re$ is the Reynolds number for a generic (peripheral) collision, varies surprisingly strongly with the magnetic field and therefore with the impact parameter $b$.

As always, we must bear in mind that holography is still far from being a precision tool; indeed one reason for pursuing studies such as the present one is to determine how seriously holographic predictions regarding the behaviour of the QGP should be taken. In the present case, there is reason to doubt the validity of the precise numerical predictions of the model: the reader may prefer to interpret our results as simply implying that holography indicates that there might be a significant difference between the Reynolds number of the plasma produced by a nearly central collision and that of its counterparts appearing in peripheral collisions.

Nevertheless, even at this qualitative level, one might think that the rather dramatic predicted effect of magnetic fields on the Reynolds number or kinematic viscosity would be easily detected in the data from experimental studies focusing on centrality dependence \cite{kn:centrality1,kn:centrality2}. But this may not be the case. For while our (partly conjectural) Figure 4 does show a strong variation of $\nu/\nu_0$ with $b$, most of the variation occurs in the region of very small $b$, the graph being otherwise rather flat; depending on the resolution, this could mean that the effect will appear to be independent of centrality. (On the other hand, it should be sensitive to global parameters, such as the impact energy.)

Furthermore, the extreme magnetic fields we have been considering, and of course consequently their influence on $\nu/\nu_0$ and $Re$, are thought to be strongly time-dependent, and it is not yet clear whether they persist for a sufficiently long time to have a significant observable effect. It will be interesting to see whether variations of the Reynolds number can be detected in future studies, either directly or perhaps indirectly; for example, in investigations of paramagnetic squeezing \cite{kn:bali,kn:gergend}.

\addtocounter{section}{1}
\section*{\large{\textsf{Acknowledgements}}}
The author is extremely grateful for the hospitality of the HECAP Section of the Abdus Salam ICTP, where this work was initiated.

\end{document}